\documentstyle[12pt]{article}

\begin{document}
\pagestyle{plain}

\newcommand{\be}{\begin{equation}}
\newcommand{\ee}{\end{equation}}
\newcommand{\bea}{\begin{eqnarray}}
\newcommand{\eea}{\end{eqnarray}}

\title{One parameter family  of  Compacton \\Solutions  in a class of
Generalized  Korteweg-DeVries Equations}
\author{Avinash Khare \\
{\small \sl Center for Nonlinear Studies, Los Alamos National
Laboratory,} \\ {\small \sl Los Alamos, NM 87545}\\
\and
{\small \sl Institute of Physics,Sachivalaya Marg} \\
{\small \sl Bhubaneswar 751005, India}
\and  Fred Cooper \\
{\small \sl Theoretical Division, Los Alamos National Laboratory,}\\
{\small \sl Los Alamos, NM 87545}}
\maketitle

\begin{abstract}
We study the generalized Korteweg-DeVries equations derivable
from the Lagrangian:
$ L(l,p) = \int \left( \frac{1}{2} \varphi_{x} \varphi_{t} - {
{(\varphi_{x})^{l}} \over
{l(l-1)}} + \alpha(\varphi_{x})^{p} (\varphi_{xx})^{2}  \right) dx, $
where the usual fields $u(x,t)$ of the generalized KdV equation
are defined by $u(x,t) = \varphi_{x}(x,t)$.
For $p$ an arbitrary continuous parameter  $0< p \leq 2
 ,l=p+2$ we find compacton solutions to these equations which
 have the feature that their width is independent of the
amplitude. This generalizes previous results which considered $p=1,2$.
For the exact compactons we find a relation between the energy, mass and
velocity of the solitons. We show
that this relationship can also be
obtained using a variational method based on the principle of least action.
\newline
\newline
PACS numbers:  03.40.Kf, 47.20.Ky, Nb, 52.35.Sb

\end{abstract}

Recently, Cooper et. al.  \cite{cooper1}  obtained compacton solutions to a
generalized sequence of KdV equations of the form:
 \begin{equation}
  u_t= u_x u^{l-2} + \alpha\left(2 u_{xxx}u^p + 4p u^{p-1} u_x
u_{xx} +p(p-1) u^{p-2} (u_x)^3 \right)\label{kstar}
\end{equation}
for the case where $l=p+2$ and $0 <p \leq 2$. In particular, they obtained
compacton solutions (i.e. solitary waves with compact support) for the cases
$p=1,2$. In \cite{cooper1}, hereafter referred to as I, it was shown that the
width of the compactons were independent of the amplitude and that all the
solutions to these equations obey the same first three conservation laws as
found in the KdV case: namely, area, mass and energy.  In I it was also
found that the energy mass and velocity of the compactons could be simply
related for the integer values of $p$ studied.  The purpose of this Comment is
to extend the work done in I to non integer values of $p$. That is we show that
for arbitrary continuous values of $p,  0 < p \leq 2$ there exist compacton
solutions to the above  equation. We obtain the explicit expressions for the
solitons and determine their mass ($M$) and energy($E$)  and  show that
\be
E= cM/(p+2) \label{eqvel}
\ee
where $c$ is the velocity of the soliton.  We also consider the class
of variational wave functions of the form:
\begin{equation}
u_{v}(x,t) = A(t) \exp \left[-\beta(t) |x-q(t)|^{2n} \right],
\label{uv} \end{equation}
where $n$ is an arbitrary continuous, real parameter. We find that this
class of trial wave functions yield the exact relationship (\ref{eqvel}) as
well as giving an excellent global approximation to the compacton except
at the endpoints.

Folowing I, we consider travelling wave solution to eq. (\ref{kstar}) of the
form
 \begin{equation}
u(x,t) =f(\xi) =f(x+ct) ,
\end{equation}
As shown in I, the function $f$ satisfies for $l=p+2$

\begin{equation}
\alpha f^{\prime 2} =  {{c} \over {2}} f^{2-p} -{{f^{2}} \over {(p+1)(p+2)}}.
\label{fp}
 \end{equation}
On using the ansatz
\be
f(\xi) = \beta  \, \rm{cos}^a(\gamma \xi)
\ee
in eq. (\ref{fp}), we then obtain the following one continuous parameter family
of compacton solutions:
\be
u=  ({ c(p+1) (p+2) \over 2 })^{1/p} {\rm{cos}}^{2/p} ({p \xi \over
\sqrt{4 \alpha (p+1) (p+2)} }) \label{exact}
\ee
where:

\be
|{p \xi \over
\sqrt{4 \alpha (p+1) (p+2)} }| \leq {\pi \over 2}
\ee

Note that for all these solutions, the width is independent of the velocity c
of the compacton. For $p=1,2$ we immediately recover the solutions obtained in
I by choosing $\alpha=1/2,3$ respectively.  The mass $M$ and Energy $E$ of
these solutions are easily calculated. We find:
\bea
M &=&{2 \over p} (2c)^{2/p} \alpha^{1/2} [(p+1) (p+2)]^{2/p+1/2}
{\Gamma^2(2/p+1/2) \over \Gamma(4/p+1)} \nonumber \\
E &=&{1 \over p(p+2)} (2c)^{2/p+1} \alpha^{1/2} [(p+1) (p+2)]^{2/p+1/2}
{\Gamma^2(2/p+1/2) \over \Gamma(4/p+1)} \label{me}
\eea
and hence the relation (2) between $E,M$ and $c$ follows immediately.

We can also study these solitary waves in the variational approximation
defined by eq. (\ref{uv}) as discussed
in I.  Following the arguments in I and using $l=p+2$ we again
discover that the soliton width $\beta$ is independent of the mass $M$ for
arbitrary $p$.  We also find that for arbitrary $p$ we exactly satisfy
the relationship (2) for any value of the variational parameter $n$.
Minimizing
the action gives us the optimum value of $n$ for each $p$ and again we find
for the optimal $n$ excellent numerical agreement for the conserved quantities
and
good agreement for the global compacton profile except near the places where
the true compacton goes to zero. As a particular example for the case
$p=4/3$ we have that the exact energy (in units of $M$) is $0.0245277$, wheras
the variational result is $0.0244173$ which is obtained at $n=1.199$. In
fig. 1 we compare the exact and variational expressions for the soliton
solutions for the case $M=1$, and $\alpha=1$.

Finally, since the generalized sequence of KdV equations of Rosenau and
Hyman \cite{RH} are very similar to those of eq. (1), we expect that a similar
one continuous parameter family of compacton solutions will also exist in their
case
when in their notation $m=n$ and $1<m \leq 3$.

\section*{Acknowledgements}
We are grateful to Philip Rosenau for useful discussions. Further, one of us
(AK) would like to thank the CNLS, Los Alamos National Labs for partial
financial support and for its hospitality.
This work was supported in part by the DOE.

\section*{FIGURE CAPTIONS}

\noindent Fig. 1: $  u_{var} $ with $n=1.199$  and $u_{exact}$ for $p=4/3$,
$M=1$ and $\alpha=1$.

\end{document}